\documentclass[superscriptaddress,onecolumn,secnumarabic,
amssymb,amsmath,nobibnotes,aps,prd,showkeys,showpacs,nofootinbib]{revtex4}

\usepackage[latin1]{inputenc}
\usepackage{graphicx}
\usepackage[english]{babel}

\usepackage{amsmath}
\usepackage{amssymb}
\usepackage{amsfonts}
\usepackage{colordvi}
\usepackage{psfrag}
\usepackage{color}

\begin{document}
\def\cL{{\cal L}}
\def\be{\begin{equation}}
\def\ee{\end{equation}}
\def\bea{\begin{eqnarray}}
\def\eea{\end{eqnarray}}
\def\beq{\begin{eqnarray}}
\def\eeq{\end{eqnarray}}
\def\tr{{\rm tr}\, }
\def\nn{\nonumber \\}
\def\e{{\rm e}}
\def\bef{\begin{figure}}
\def\eef{\end{figure}}
\newcommand{\ans}{ansatz }
\newcommand{\eeqn}{\end{eqnarray}}
\newcommand{\bd}{\begin{displaymath}}
\newcommand{\ed}{\end{displaymath}}
\newcommand{\mat}[4]{\left(\begin{array}{cc}{#1}&{#2}\\{#3}&{#4}
\end{array}\right)}
\newcommand{\matr}[9]{\left(\begin{array}{ccc}{#1}&{#2}&{#3}\\
{#4}&{#5}&{#6}\\{#7}&{#8}&{#9}\end{array}\right)}
\newcommand{\matrr}[6]{\left(\begin{array}{cc}{#1}&{#2}\\
{#3}&{#4}\\{#5}&{#6}\end{array}\right)}
\newcommand{\cvb}[3]{#1^{#2}_{#3}}
\def\lsim{\raise0.3ex\hbox{$\;<$\kern-0.75em\raise-1.1ex
e\hbox{$\sim\;$}}}
\def\gsim{\raise0.3ex\hbox{$\;>$\kern-0.75em\raise-1.1ex
\hbox{$\sim\;$}}}
\def\abs#1{\left| #1\right|}
\def\simlt{\mathrel{\lower2.5pt\vbox{\lineskip=0pt\baselineskip=0pt
           \hbox{$<$}\hbox{$\sim$}}}}
\def\simgt{\mathrel{\lower2.5pt\vbox{\lineskip=0pt\baselineskip=0pt
           \hbox{$>$}\hbox{$\sim$}}}}
\def\unity{{\hbox{1\kern-.8mm l}}}
\newcommand{\eps}{\varepsilon}
\def\ep{\epsilon}
\def\ga{\gamma}
\def\Ga{\Gamma}
\def\om{\omega}
\def\omp{{\omega^\prime}}
\def\Om{\Omega}
\def\la{\lambda}
\def\La{\Lambda}
\def\al{\alpha}
\newcommand{\ov}{\overline}
\renewcommand{\to}{\rightarrow}
\renewcommand{\vec}[1]{\mathbf{#1}}
\newcommand{\vect}[1]{\mbox{\boldmath$#1$}}
\def\tm{{\widetilde{m}}}
\def\mcirc{{\stackrel{o}{m}}}
\newcommand{\Dm}{\Delta m}
\newcommand{\dm}{\varepsilon}
\newcommand{\tanb}{\tan\beta}
\newcommand{\nbar}{\tilde{n}}
\newcommand\PM[1]{\begin{pmatrix}#1\end{pmatrix}}
\newcommand{\up}{\uparrow}
\newcommand{\down}{\downarrow}
\def\omE{\omega_{\rm Ter}}
%

\newcommand{\Dsusy}{{susy \hspace{-9.4pt} \slash}\;}
\newcommand{\DCP}{{CP \hspace{-7.4pt} \slash}\;}
\newcommand{\mc}{\mathcal}
\newcommand{\gr}{\mathbf}
\renewcommand{\to}{\rightarrow}
\newcommand{\gtc}{\mathfrak}
\newcommand{\wh}{\widehat}
\newcommand{\br}{\langle}
\newcommand{\kt}{\rangle}


\def\lsim{\mathrel{\mathop  {\hbox{\lower0.5ex\hbox{$\sim$}
\kern-0.8em\lower-0.7ex\hbox{$<$}}}}}
\def\gsim{\mathrel{\mathop  {\hbox{\lower0.5ex\hbox{$\sim$}
\kern-0.8em\lower-0.7ex\hbox{$>$}}}}}

\def\nn{\\  \nonumber}
\def\de{\partial}
\def\brf{{\mathbf f}}
\def\bbf{\bar{\bf f}}
\def\bF{{\bf F}}
\def\bbF{\bar{\bf F}}
\def\bA{{\mathbf A}}
\def\bB{{\mathbf B}}
\def\bG{{\mathbf G}}
\def\bI{{\mathbf I}}
\def\bM{{\mathbf M}}
\def\bY{{\mathbf Y}}
\def\bX{{\mathbf X}}
\def\bS{{\mathbf S}}
\def\bb{{\mathbf b}}
\def\bh{{\mathbf h}}
\def\bg{{\mathbf g}}
\def\bla{{\mathbf \la}}
\def\bmu{\mathbf m }
\def\by{{\mathbf y}}
\def\bmu{\mbox{\boldmath $\mu$} }
\def\bsig{\mbox{\boldmath $\sigma$} }
\def\bunity{{\mathbf 1}}
\def\cA{{\cal A}}
\def\cB{{\cal B}}
\def\cC{{\cal C}}
\def\cD{{\cal D}}
\def\cF{{\cal F}}
\def\cG{{\cal G}}
\def\cH{{\cal H}}
\def\cI{{\cal I}}
\def\cL{{\cal L}}
\def\cN{{\cal N}}
\def\cM{{\cal M}}
\def\cO{{\cal O}}
\def\cR{{\cal R}}
\def\cS{{\cal S}}
\def\cT{{\cal T}}
\def\eV{{\rm eV}}

\title{ Chaotic Solutions and Black Hole Shadow in $f(R)$ gravity}

\author{Andrea Addazi}

\affiliation{ Center for Theoretical Physics, College of Physics Science and Technology, Sichuan University, 610065 Chengdu, China}
\affiliation{INFN sezione Roma {\it Tor Vergata}, I-00133 Rome, Italy, EU}

\author{Salvatore Capozziello}

\affiliation{Dipartimento di Fisica, {\it E. Pancini} Universita {\it Federico II} di Napoli,
Compl. Univ. Monte S. Angelo Ed. G, Via Cinthia, I-80126 Napoli (Italy)}

\affiliation{INFN Sez. di Napoli, Compl. Univ. Monte S. Angelo Ed. G, Via Cinthia, I-80126 Napoli, Italy}

\affiliation{Scuola Superiore Meridionale, Largo S. Marcellino 10, I-80138 Napoli, Italy}

\affiliation{Laboratory for Theoretical Cosmology, Tomsk State University of Control Systems and Radioelectronics (TUSUR), 634050 Tomsk (Russia)}

\author{Sergei Odintsov}

\affiliation{Institute of Space Sciences (IEEC-CSIC) C. Can Magrans s/n, 08193 Barcelona, Spain}

\affiliation{ICREA, Passeig Luis Companys, 23, 08010 Barcelona, Spain}

\date{\today}

\begin{abstract}
We discuss the emergence of  black hole shadow and  photon-sphere in the context of $f(R)$ gravity.
It is shown that the shadow is exponentially sensitive to  linear instabilities of metric coming from  some $f(R)$ solutions.  Thus, the instabilities of  photon circular trajectories, delimiting the black hole photon-sphere, 
are double exponentialized. Specifically we individuate two Lyapunov exponents, rather than only one, related to 
two different sources of chaos in geodesic orbits as a sort of  {\it butterfly effect}. Such a result violates 
the black hole chaos bound proposed by {\it Maldacena}, {\it Shenker} and {\it Stanford} for General Relativity. 
We also explore the impact of the black hole metric instabilities in $f(R)$ gravity on the
 quasi-normal modes.  In the framework of Extended Theories of Gravity, our analysis suggests a new paradigm  to deal with  black hole shadow  and gravitational waves observations coming from black hole merging in the ringdown phase. 
\end{abstract}

\maketitle

\section{Introduction}
One of the most fascinating aspects of black hole  (BH) physics
is that, below a critical distance, 
also light gets trapped in circular orbits.
In other words, no any signal can escape from the BH gravitational attraction
below a radial threshold. 
The photons orbit around the "BH design" giving rise to the so dubbed BH {\it photon-sphere}.
Indeed, the BH appears as a {\it black hole shadow} 
for any distant observers, surrounded by a luminous photon ring. 
In General Relativity (GR), the photon-sphere is limited by 
a typical radius $r_{C}$ which is related to the Schwarzschild radius $r_{S}$
as $2r_{C}=3r_{S}$ \cite{f,fi}. 
Such a circular null-like geodesic 
is exponentially unstable against arbitrary small geodesic perturbations, 
as a manifestation of the  so-called {\it butterfly effect}.
In other words,  here there is the emergence of {\it chaos}:
any deviation from the critical photon-sphere trajectory 
leads to instabilities, exponentially growing in time. 
Chaos is ubiquitously appearing in many different aspects of Nature, 
notoriously including atmospheric physics, finance, biology and so on.  Thus, it is not surprising to 
find "chaos footprints" in BH physics. 
Specifically, a Lyapunov exponent, related to the BH shadow, parametrically controls the exponential chaos growth in time.
Maldacena,  Shenker, and  Stanford (MSS) conjectured that 
such instabilities cannot grow more than exponentially, with a 
 Lyapunov exponent which cannot exceed a critical value directly proportional the {\it Bekeinstein-Hawking} temperature \cite{Maldacena:2015waa}\footnote{It is worth mentioning that recently  Bianchi et al. showed that in the case of {\it fuzzballs}, 
the Lyapunov chaotic exponent deviates, but never exceeds, from the MSS bound \cite{Bianchi:2020des}.}. 
In general, chaos can be "diagnosed" from analyzing 
the out-of-time-order (OTO) correlators
related to the commutators of time-separated operators \cite{Maldacena:2015waa}. 
In chaotic systems, the OTO Operators have exponential instabilities controlled by the Lyapunov exponent\footnote{The possibility that
chaos may be also relevant in the BH interior and information processing 
was explored in Refs.\cite{Addazi:2015gna,Addazi:2016cad,Addazi:2015hpa} .}. 
It is also interesting to note that the circular unstable photon-sphere 
is related to the characteristic modes of BH -- the so-called quasi-normal modes (QNMs). 
More precisely, the BH Lyapunov exponent is proportional to the QNM decays 
after the BH formation. This suggests that BH shadow issues are also related to possible 
tests of the BH horizon from gravitational waves emitted from the BH merging during the ring-down phase,
when a new BH has formed undergoing to relaxation.
It is worth  stressing how QNMs are crucially important for the  gravitational waves physics after 
LIGO/VIRGO observations \cite{LIGO1,LIGO2,LIGO3,LIGO4}. 

\vspace{0.1cm}

On the other hand, the above considerations  may be significantly improved
considering  Extended Theories of Gravity \cite{f4}. In principle, any extensions of GR may 
lead to revisit our standard conceptions of BH physics, including stability issues 
and BH thermodynamics.  
The possibility that the standard Einstein-Hilbert (EH) action may be 
embedded and corrected   extending the gravitational action considering further curvature (and torsion invariants), besides the Ricci curvature scalar $R$,
 is one of the most debated issue of contemporary 
gravitational physics and cosmology. 
Indeed, within the vast landscape of possible  theories of gravity, which are theoretically self-consistent and  pass 
astrophysical and cosmological tests (specifically the Solar System ones), there are several  models leading to new  predictions and solutions in BH physics.  

First of all, the Birkhoff theorem for  spherically symmetric BHs can be violated 
in many gravity theories beyond EH. Let us remind that,  
in GR, the Birkhoff theorem states that any spherically symmetric solution of Einstein field equations is  static and stationary.  
Despite of this statement, many spherically symmetric solutions of Extended Gravity 
are unstable under linear metric perturbations \cite{BH,NO0,NO,NO2,Addazi:2016prb,Addazi:2016hip,Addazi:2017puj,Addazi:2017lat,Addazi:2017vti,Addazi:2017cim,Sebastiani:2013fsa}.
Linear metric instabilities were found in many theories, including $f(R)$ gravity, for extremal Schwarzschild-de Sitter and Reissner-Nordstrom BHs \cite{Addazi:2018pcc, Diego}. 
Furthermore, several BH solutions have been found and studied in the context of $f(R)$ gravity, see \cite{Cruz1,Cruz2, RCai, Clifton, Faraoni1, Faraoni2, Sebastiani1, Olmo}.
\vspace{0.1cm}

In this paper, we are going to analyze the BH shadow, the chaos photon-sphere instabilities and the QNMs 
in $f(R)$ gravity. 
We find that the OTO correlators 
are double exponentially sensitive for two
characteristic Lyapunov exponents rather than the only  photon-sphere one. 
Such a result violates the MSS bound. 
On the other hand, the scalaron field instabilities, related to $f(R)$,  must be  fine-tuned 
against an exponential function of time.  If not so, 
an obvious contradiction with lensing and gravitational waves data 
would rule out $f(R)$ gravity and similar  models that seem retained from observations \cite{Lombrisier}
(see  also Ref.\cite{Addazi:2018pcc} for an analysis of Active Galactic Nuclei  and lensing data). 

This approach  may suggest several  opportunities to tests  $f(R)$ gravity and, in general, Extended Theories of Gravity, by BH physics.  First of all, the observations by the  the Event Horizon Telescope (EHT)  collaboration of the ${\rm M87^{*}}$ BH shadow \cite{EHT1,EHT2,EHT3} can be used to probe possible anomalous growth or decrease of the BH area.
In general, any supermassive compact object can be adopted to test gravity and forthcoming observations towards our Galactic center Sgr A$^{*}$, as recently reported in \cite{Fromm}, will constitute a formidable laboratory in this perspective.

Specifically, $f(R)$ gravity can predict BHs with an event horizon  increasing in radius 
in time.  Furthermore horizonless solutions can be achieved in this context \cite{WH}. 
Finally, the new Lyapunov exponent emerging in $f(R)$ gravity is related to the quasi normal modes (QNMs) of BH in the ringdown phase.
Thus,  metric instabilities coming from  $f(R)$ gravity, beyond the Einstein theory, can be, in principle,  
tested, and then confirmed or rejected, from gravitational waves observations after the BH or neutron star merging events to a final BH state. This phenomenology could be of
high interest for LIGO/VIRGO \cite{LIGO1,LIGO2,LIGO3,LIGO4} and KAGRA \cite{Akutsu:2020zlw} collaborations.

\vspace{0.5cm}

To fix the ideas, let us  consider OTO correlators and BH chaos: 
the classical chaos growing is related to an exponential instability in
correlators of particle trajectory coordinates, 
i.e. for example, the azimuthal angles $\varphi\equiv\varphi(t)$: 
\begin{equation}
\label{eq2}
C(t',t)\equiv\Big\{\varphi(t'),\varphi(t)\Big\}_{P.B.}=\frac{\delta\varphi(t)}{P_{\varphi}}= c\, e^{\lambda(t-t')}
\end{equation}
with  $c=C(t=0)$, $\{... \, , ...\}_{P.B.}$ the Poisson brackets, $P_{\varphi}$ the conjugate momentum, $t$ the time variable
and $t'$ the initial fixed time. 
The constant $\lambda$  introduced above is the Lyapunov exponent,
which controls the chaos instability of circular photon trajectories. 
Eq.\ref{eq2} can be generalized to a quantum version 
for generic correlators as follows 
\begin{equation}
\label{kQNMM4ajaj}
\hat{C}(t',t)=-\langle [\hat{W}(t),\hat{V}(t')]^{2}\rangle\, , 
\end{equation}
where $\hat{W},\hat{V}$ are two generic Hermitian correlators in the Heisenberg representation, 
$\langle ... \rangle=Z^{-1}{\rm Tr}[e^{-\beta \hat{H}}]$\, 
is the thermal expectation value at temperature $T=\beta^{-1}$. $Z$ and $H$ are the partition function and the Hamiltonian of the N-body system respectively. 
Eq.\eqref{eq2} indicates that small deviations from the initial conditions
related to trajectories diverge as $e^{\lambda t}$ (conventionally $t'=0$)
with a characteristic time constant 
$\tau \sim \lambda ^{-1}$. The $C$ in Eq.\eqref{eq2} and $\hat{C}$ in Eq.\eqref{kQNMM4ajaj} start to be close to one after a critical "scrambling time" $t_{*}\sim \lambda^{-1}\log c^{-1}$. 

According to MSS, any correlator in N body systems has 
a Lyapunov exponent that is below a {\it critical thermal Lyapunov exponent}: 
\begin{equation}
\label{kQNMM5}
C(t)\equiv C(t'=0,t)=c\,  {\rm exp} \lambda\, t+....,\,\,\,\, {\rm with}\,\,\,\, \lambda\leq \lambda_{T}=2\pi k_{B}T/\hbar\, . 
\end{equation}
In the case of a BH, the maximal Lyapunov exponent 
corresponds to a $\lambda_{BH}=2\pi k_{B}T_{BH}/\hbar$ where $T_{BH}$ is the Bekeinstein-Hawking temperature
and $\hbar,k_{B}$ are the Planck and Boltzmann constants,  respectively.

Eqs.\eqref{eq2} indicates that the the chaotic instabilties are 
exponentially sensitive to the Lyapunov exponent, in turn 
related to the BH radius linear variations: 
\begin{equation}
\label{eq3}
r_{BH}+\delta r_{BH}\rightarrow \lambda+\delta \lambda\,. 
\end{equation}
This leads to the relative relations among the variations of the Lyapunov exponent, the BH temperature and the BH radius as follows: 
\begin{equation}
\label{eq4}
\frac{\delta \lambda}{\lambda}=\frac{\delta T}{T}=\frac{\delta r_{BH}}{r_{BH}}\, 
\end{equation}
while the correlators relative to the linear variations are related to the temperature as follows: 
\begin{equation}
\label{eq5}
\frac{\delta C}{C}=t \, \delta \lambda=(2\pi k_{B}/\hbar) t  \,\delta T\, . 
\end{equation}
In a standard BH, in (semi)classical GR, 
if infalling matter/energy as well as Hawking radiation are negligible,
the  Lyapunov BH coefficient is constant in time, i.e. the expectation value  is $\langle \delta \lambda(t) \rangle=0$.
However, if the Lyapunov exponent has a time dependence, the 
linear correlator variation has a non-linear dependence inside the exponential function of time. 
Such a phenomenon can be extremely interesting in Extended Theories of Gravity, where there is a class of spherically symmetric solutions 
which are unstable under metric perturbations \cite{f4,NO,NO2,Addazi:2016prb,Addazi:2016hip,Addazi:2017puj,Addazi:2017lat,Addazi:2017vti,Addazi:2017cim,Sebastiani:2013fsa,Addazi:2018pcc}. 
Such a dynamics is  a consequence of the energy conditions where
 the presence of new degrees of freedom, due to Extended Gravity,
play an important role into dynamics \cite{Addazi:2017vti, Lobo}. 

This fact gives rise to a variation in time of the BH radius, 
in turn altering the thermal Lyapunov BH exponent in time: 
\begin{equation}
\label{LLL}
\langle \delta \lambda_{BH}(t)\rangle =\lambda_{BH}(0)[{\rm exp}(\lambda_{g} t)-1]\, ,
\end{equation}
where $\lambda_{g}$ is a new Lyapunov exponent related to the metric instabilities\footnote{In the following discussions we will often omit the average $\langle ... \rangle$ notation for a lighter notation.}.  

Thus, the maximally critical correlator has a double exponential form as 
\begin{equation}
\label{akak}
C_{critical}(t)=\epsilon \, e^{t\lambda_{BH}^{0}e^{\lambda_{g}t}}\simeq e^{\lambda_{BH}^{0}t+\lambda_{BH}^{0}\lambda_{g}t^{2}/2+...}\, ,
\end{equation}
where $\lambda_{BH}^{0}=\lambda_{BH}(0)$. 
The $O(t^{2})$ and $O(t)$ terms in the exponential of Eq.\eqref{akak}
are comparable when the time is $t\simeq \lambda_{g}^{-1}$. 
In the case $\lambda_{g}<<\lambda_{BH}$, the double exponentialization is completely negligible compared to the leading MSS exponential
and Eq.\eqref{eq2} is re-obtained. However, if $\lambda_{g}\simeq \lambda_{BH}$
the metric instability cannot be neglected anymore in the chaos correlator
after a scrambling time.

In the next sections, we will discuss  the instabilties in $f(R)$ gravity 
and their implications in chaos around the BH shadows and QNMs.

\section{Chaotic instabilities in $f(R)$ gravity}

In this section, we will describe the metric instability phenomena in  
$f(R)$ gravity (see Refs.\cite{f2,f3,f4,f5} for detailed  reviews on the subject). 

$f(R)$ gravity  is a straightforward extension of the Einstein-Hilbert action  where also non-linear terms in the Ricci scalar can be included. It is: 
\begin{equation}
\label{action}
I=\frac{1}{16\pi}\int d^{4}x\sqrt{-g}f(R)+S_{M}
\end{equation}
where $R$ is the Ricci scalar curvature, $g\equiv {\rm det}g_{\mu\nu}$ is the determinant of the metric tensor. Specifically,
$f(R)$ is a generic function of $R$ and  $S_{M}$ is the matter action. We set the Newton constant to $G_{N}=1$. 
The field equations read as 
\begin{equation}
\label{EoM}
f_{R}(R)R_{\mu\nu}-\frac{1}{2}f(R)g_{\mu\nu}-[\nabla_{\mu}\nabla_{\nu}-g_{\mu\nu}\Box]f_{R}(R)=8\pi T_{\mu\nu}
\end{equation}
where $T_{\mu\nu}$ is the stress-energy tensor of matter and $f_{R}(R)\equiv df/dR$. 
As it is well known, Eq.\eqref{EoM} admits spherically symmetric metric solutions \cite{Stabile} of the form 
\begin{equation}
\label{eqg1}
ds^{2}=f(r)dt^{2}-\frac{1}{g(r)}dr^{2}-r^{2}d\Omega^{2}\, ,
\end{equation}
where it is possible to restrict to the sub-class of solutions
\begin{equation}
\label{dS}
f(r)=g(r)=1-\frac{2M}{r}-\Big(\frac{r}{L_{dS}}\Big)^{2}\, , 
\end{equation}
where $M$ is the BH mass and $L_{dS}^{2}=3/\Lambda$ is the de Sitter curvature radius. Here
$\Lambda$ is the cosmological constant. 
It is worth noticing that a particular case of Eq.\eqref{eqg1} is the Schwarzschild-de Sitter solution, corresponding to $f,g$ reported in Eq.\eqref{dS}. As discussed in \cite{Stabile}, any $f(R)\neq R$ model in vacuum can be reduced to $R+\Lambda$.
Thus $f,g$ can be rewritten as 
\begin{equation}
\label{ff}
f=g=\frac{1}{L_{dS}^{2}r}(r-r_{+})(r_{\Lambda}-r)(r-r_{0})\, ,
\end{equation}
where $r_{0}=-r_{+}-r_{\Lambda}$. Here $r_{+}$ and  $r_{\Lambda}$ are the BH and cosmological horizon 
respectively.
The two radii are related  with the dS-length and the BH mass as follows 
\begin{equation}
\label{rel1}
r_{+}^{2}+r_{+}r_{\Lambda}+r_{\Lambda}^{2}=L_{dS}^{2}\, , 
\end{equation}
\begin{equation}
\label{rel2}
r_{+}r_{\Lambda}(r_{+}+r_{\Lambda})=2ML_{dS}^{2}\, . 
\end{equation}
Let us consider a near-extremal limit where the BH radius is almost saturating the Hubble radius: 
\begin{equation}
\label{ne}
r_{\Lambda}-r_{+}<<r_{+}\, . 
\end{equation}
In this limit, it is
\begin{equation}
\label{ne1}
r_{0}\simeq -2r_{+}^{2},\,\,\, L_{dS}\simeq \sqrt{3}r_{+},\,\,\, M\simeq \frac{r_{+}}{3}\, ,
\end{equation}
so that all quantities can be expressed in terms of $r_+$. The null-geodesic circular orbit has a radius
\begin{equation}
\label{ne2}
r_{C}=\frac{3}{2}\Big(1-\frac{r_{+}^{2}}{L_{dS}^{2}} \Big)r_{+}\, , 
\end{equation}
associated to a circular orbit angular velocity 
\begin{equation}
\label{ne3}
\Omega_{C}=\frac{r_{\Lambda}-r_{+}}{2r_{+}^{2}} . 
\end{equation}
In the near-extremal limit, one obtains the Nariai BH metric solution, that can be written as follows: 
\begin{equation}
\label{N1}
ds^{2}=e^{2\rho}(-d\tau^{2}+dx^{2})+e^{-2\phi}d\Omega^{2}\,,
\end{equation}
being 
\begin{equation}
\label{N2}
e^{2\rho}=-\frac{1}{\Lambda \cos^{2}\tau},\,\,\ e^{-2\phi}=\frac{1}{\Lambda}\,\,\,, \tau={\rm arcos}[\cosh\, t]^{-1}\, ,
\end{equation}
where $-\infty<t<\infty$ corresponds to $-\pi/2<\tau<\pi/2$. 
Eqs.\eqref{EoM} in the background \eqref{N1}  read as:

$$0=-\frac{e^{2\rho}}{2}f(R)-(-\ddot{\rho}+2\ddot{\phi}+\rho''-2\dot{\phi}^{2}-2\rho'\phi'-2\dot{\rho}\dot{\phi})f_{R}+\ddot{f}_{R}-\dot{\rho}\dot{f_{R}}-\rho'f'_{R}$$
\begin{equation}
\label{eqM1}
+e^{2\phi}\Big( -\frac{\partial}{\partial \tau}\Big(e^{-2\phi}\dot{f}_{R}\Big)+\frac{\partial}{\partial x}\Big(e^{-2\phi}f'_{R}\Big)\Big)\, , 
\end{equation}
$$0=\frac{e^{2\rho}}{2}f(R)-(\ddot{\rho}+2\phi''-\rho''-2\phi'^{2}-2\rho'\phi'-2\dot{\rho}\dot{\phi})f_{R}+f''_{R}-\dot{\rho}\dot{f}_{R}-\rho'f'_{R}$$
\begin{equation}
\label{eqM2}
-e^{2\phi}\Big( -\frac{\partial}{\partial \tau}\Big(e^{-2\phi}\dot{f}_{R}\Big)+\frac{\partial}{\partial x}\Big(e^{-2\phi}f'_{R}\Big)\Big)\, , 
\end{equation}
\begin{equation}
\label{eqM3}
0=-(2\dot{\phi}'-2\phi'\dot{\phi}-2\rho'\dot{\phi}-2\dot{\rho}\phi')f_{R}+\dot{f}'_{R}-\dot{\rho}f'_{R}-\rho'\dot{f}_{R}\, , 
\end{equation}
$$0=\frac{e^{-2\phi}}{2}f(R)-e^{-2(\phi+\rho)}(-\ddot{\phi}+\phi''-2\phi'^{2}+2\dot{\phi}^{2})f_{R}-f_{R}+e^{-(\rho+\phi)}\Big(\dot{\phi}\dot{f}_{R}-\rho'f_{R}'\Big)$$
\begin{equation}
\label{eqM4}
-e^{-2\rho}\Big( -\frac{\partial}{\partial \tau}\Big(e^{-2\phi}\dot{f}_{R}\Big)+\frac{\partial}{\partial x}\Big(e^{-2\phi}f'_{R}\Big)\Big)\, , 
\end{equation}
with $A'\equiv \partial A/\partial x$ and $\dot{A}\equiv \partial A/\partial \tau$. 
Eqs.\eqref{eqM1}, \eqref{eqM2}, \eqref{eqM3}, \eqref{eqM4} correspond to the $\{0,0\},\{1,1\},\{0,1\}$ and $\{2,2\}$,$\{3,3\}$ components of  Eqs.\eqref{EoM} in  Eq.\eqref{N1} background,
respectively. 

The perturbations of the Nariai background can be written 
as
\begin{equation}
\label{rhoa}
\rho=-{\rm ln}[\sqrt{\Lambda}\cos \tau]+\delta \rho,\,\,\, \phi={\rm ln}\sqrt{\Lambda}+\delta \phi\, . 
\end{equation}
The perturbation equations are obtained substituting Eqs.\eqref{rhoa}
into Eqs.\eqref{eqM1}, \eqref{eqM2}, \eqref{eqM3}, \eqref{eqM4}
and they correspond to
$$0=\frac{-f_{R}(R_{0})+2\Lambda f_{RR}(R_{0})}{2\Lambda \cos^{2}\tau}\delta R-\frac{f(R_{0})}{\Lambda \cos^{2}\tau}\delta \rho-f_{R}(R_{0})(-\delta \ddot{\rho}+2\delta \ddot{\phi}+\delta \rho''-2\tan \tau \delta \dot{\phi}) $$
\begin{equation}
\label{pert1}
-\tan \tau f_{RR}(R_{0})\delta \dot{R}+f_{RR}(R_{0})\delta R''\, , 
\end{equation}
$$0=-\frac{-f_{R}(R_{0})+2\Lambda f_{RR}(R_{0})}{2\Lambda \cos^{2}\tau}\delta R+\frac{f(R_{0})}{\Lambda \cos^{2}\tau}\delta \rho
-f_{R}(R_{0})(\delta \ddot{\rho}+2\delta \phi''-\delta \rho''-2\tan \tau\delta \dot{\phi})$$
\begin{equation}
\label{pert2}
+f_{RR}(R_{0})\delta \ddot{R}-\tan \tau f_{RR}(R_{0})\delta \dot{R}\, , 
\end{equation}
\begin{equation}
\label{pert3}
0=-2f_{R}(R_{0})(\delta \dot{\phi}'-\tan \tau \delta \phi')+f_{RR}(R_{0})(\delta \dot{R}'-\tan \tau\delta R')\, , 
\end{equation}
$$0=- \frac{-f_{R}(R_{0})+2\Lambda f_{RR}(R_{0})}{2\Lambda}\delta R-\frac{f(R_{0})}{\Lambda}\delta \phi-\cos^{2}\tau f_{R}(R_{0})(-\delta \ddot{\phi}+\delta \phi'' ) $$
\begin{equation}
\label{pert4}
-\cos^{2}\tau f_{RR}(R_{0})(-\delta \ddot{R}+\delta R'')\, ,
\end{equation}
where 
\begin{equation}
\label{RR}
\delta R=4\Lambda(-\delta \rho+\delta \phi)+\Lambda \cos^{2}\tau(2\delta \ddot{\rho}-2\delta \rho''-4\delta \ddot{\phi}+4\phi'')\, . 
\end{equation}
As it is well known, this set of linear equations contain unstable modes 
as follows (see Ref.\cite{Sebastiani:2013fsa}): 

\begin{equation}
\label{kka}
\delta \phi(t)=\phi_{0}e^{a_{\pm}t},\,\,\, a_{\pm}=\frac{-1\pm \sqrt{1+4m^{2}}}{2}, \,\,\, r_{+}=\frac{e^{-\phi}}{\sqrt{\Lambda}}\, . 
\end{equation}
where $t$ is the cosmological time (not $\tau$).
Here $m$ is the normalized mass in $G_{N}=1$ dimensions. 
The stability is related to the following parameter
\begin{equation}
\label{aaa}
\alpha=\frac{2\Lambda f_{RR}(R_{0})}{f'(R_{0})}
\end{equation}
in turn related to Eqs.\eqref{kka} and the following linear equation: 
\begin{equation}
\label{stability}
\frac{d^{2}\delta \phi}{dt^2}+\tanh t\frac{d\delta \phi}{dt}-m^{2}\delta \phi=0,\,\,\mbox{with}\, \,m^{2}=\frac{2(2\alpha-1)}{3\alpha}\, , 
\end{equation}
 which simplifies in the regime $\tanh t\simeq 1$ for $t>>0$. In this situation, $0<\alpha<1/2$ or $\alpha<0$ are the stability regions, while there are two new branches related to the Lyapunov exponents: 
\begin{equation}
\label{lalal}
\lambda_{g\pm }\equiv \frac{-1\pm \sqrt{1+4m^{2}}}{2}\,.
\end{equation}
In the case of solution  $\lambda_{g+}$, a critical branch  is comparable to the  Lyapunov BH exponent. 
It corresponds to  compare  the linear perturbation effective mass and the BH temperature, that is 
\begin{equation}
\label{BBB}
\lambda_{BH}=\lambda_{g+} \rightarrow m^{2}\sim T\sim 1/r_{+},\,\,\,\, {\rm if}\,\,\,m<<1\,.
\end{equation}
Here $m^{2}$ and $T$ are both adimensional being $G_{N}=M_{Pl}=c=\hbar=k_{B}=1$.
Thus, for macroscopic BHs, the effective (adimensional) mass for the scalar metric perturbation must be extremely suppressed in order to not have a double exponentiation in chaos correlators. 
In other words, if $m^{2}>2\pi T$,  one would expect that the {\it MSS bound is violated}:
there is a dominant Lyapunov exponent {\it not} related to the Hawking temperature
and corresponding to the linear metric instability. 
In Fig.2, some BH Lyapunov branches are shown in the cases of  evaporating and antievaporating solutions. 

An important issue   is if  realistic classes of $f(R)$gravity models exist exhibiting such a phenomenon
or if these aspects are confined within  unphysical toy models. 
Interestingly, the metric instabilities are present in many   $f(R)$ gravity models. In this perspective 
Let us consider three $f(R)$ cases interesting for  cosmology and relativistic astrophysics. 

\subsection{ The polynomial model} 
This class of models have been largely investigated starting from the early $f(R)=R+\gamma_2 R^2$. It can be easily generalized as
\begin{equation}
\label{pol}
f(R)=R+\gamma_{n} R^{n},\,\,\, n>0\, ,
\end{equation}
where $\gamma_{n}$ is a constant with dimensionality dependent 
on the power $n$. 
In this case, the de Sitter solution is 
obtained for 
\begin{equation}
\label{RR}
R=R_{0}=4\Lambda=\Big(\frac{1}{\gamma_{n}(n-2)} \Big)^{\frac{1}{n-1}}\,,\,\,\,\, n\neq 2\, .
\end{equation}
In the dS-branch, Eq.\eqref{aaa} corresponds to 
\begin{equation}
\label{alp}
\alpha=\frac{n-1}{2\Big(1+\frac{n-2}{n}\Big)}\, . 
\end{equation}
It is worth  noticing that the stability conditions 
from Eq.\eqref{stability} are violated for any 
$n>2,\gamma_{n}>0$. Interestingly, 
$n=2$ lives on the edge of the stability condition
and not any metric instabilties are predicted, being
$\alpha=1/2$ and $n^{2}=0$.

\subsection{ The exponential model} 
Another interesting case is given by exponential model, see Refs. \cite{exp1,exp2,exp3}. The Lagrangian is
\begin{equation}
\label{esp}
f(R)=R-2\bar{\Lambda} [1-e^{-R/\bar{\Lambda}}]\, .
\end{equation}
Such a model can easily converges to the $\Lambda$CDM model  
 while  all Solar System tests are recovered in the post-Newtonian limit.
In this case, a de-Sitter solution is obtained for
$R_{0}=4\Lambda=3.75\Lambda_{eff}$, where $\Lambda\simeq 0.95 \Lambda_{eff}$. 
The corresponding $\alpha=0.09$ is compatible with 
the stability bounds; i.e. in the case of the exponential model
the metric instabilities are turned off.

\subsection{ The Hu-Sawicki model} 
This model is particularly interesting to explain the accelerated behavior of cosmic fluid in the framework of $f(R)$ gravity, see \cite{HS}. The Lagrangian is 
\begin{equation}
\label{HS}
f(R)=R-2\bar{\Lambda}\Big[1-\frac{\bar{\Lambda}^{4}}{R^{4}+\bar{\Lambda}^{4}}\Big]\, . 
\end{equation}
A de Sitter solution is achieved for $R_{0}=4\Lambda$, where
for $R_{0}=3.95\bar{\Lambda}$ and $\Lambda=0.99\bar{\Lambda}$. 
The corresponding $\alpha= 0.02$ lies into the 
metric stability condition.

\vspace{0.1cm}

To summarize, the metric instabilties propagate in  polynomial 
$f(R)$ models while they are suppressed in  exponential and Hu-Sawicki models.

\section{Chaos and the Maldacena-Shenker-Stanford  conjecture}

In the previous section, we showed the presence of a new Lyapunov exponent, related to the $f(R)$ action, which can be larger than the MSS bound. 
Here, we wish to explore and clarify how the MSS violation can be possible in $f(R)$ gravity by analyzing generic Out of Thermal Equilibrium (OTO) correlators.

\vspace{0.1cm}

As mentioned in Introduction, the MSS conjecture states that, considering a generic N-particle system, the influence of chaos in commutators of two (Hermitian) operators
can evolve in time {\it no faster than exponentially} and with a Lyapunov exponent $\lambda\leq 2\pi T$ $(k_{B}=\hbar=1)$, where $T$ is the temperature of the system. In the case of BHs, the temperature coincides with the Hawking temperature. 
Eq.\eqref{kQNMM4ajaj} is an example of operator which provides a measure of chaos propagation in the N-field system.
In particular, if such a correlator can be effectively factorized in form
$C(t)\simeq 2\langle V(0)V(0)\rangle \langle  W(t)W(t)\rangle$ for large $t$, 
then this will be a clear manifestation of the {\it butterfly effect}\footnote{In this section, the reference time $t'$, appearing in Eq.\eqref{kQNMM4ajaj}, is set to zero,
as $C(t)=C(t',t)|_{t'=0}$, without losing the generality of the arguments.}. 

In the time behavior  of $C$,  there are two important characteristic time scales:
the dissipation time or collision time $t_{d}$, which is  typically $t_{d}\sim \beta=1/T$ in strongly coupled systems,
which controls 
the exponential decay of the two point function $\langle V(0)V(0)\rangle$
and 
$\langle V(0)V(0)W(t)W(t)\rangle\sim \langle V(0)V(0)\rangle \langle W(t)W(t)\rangle+O(e^{-t/t_{d}})$;
the {\it scrambling time} $t_{*}$ satisfying $C(t_{*})\sim O(1)$, 
i.e. for $C(t)\sim \hbar^{2}e^{\lambda t}$, it corresponds to $t_{*}\sim \lambda^{-1}{\rm log}\, \hbar^{-1}$.  
Typically, in the macroscopic limit, $t_{d}<<t_{*}$,  their ratio is proportional to the Planck constant 
$\hbar$ $(\rightarrow 0$ in the classical limit). 

While the C-operator can be safely computed in condensed matter lattice systems,
it  has regularization problems in Quantum Field Theory as reported in \cite{Maldacena:2015waa}.
Alternatively, a  more controllable prescription consists in treating the
following correlator: 
\begin{equation}
\label{new}
\tilde{C}(t)=-{\rm Tr}[\zeta^{2}[W(t),V(0)]\zeta^{2}[W(t),V(0)]],\,\,\,\,\,\,\, \zeta^{4}=\frac{1}{Z}e^{-\beta H}\,.
\end{equation} 
Such a correlator is related to 
\begin{equation}
\label{coc}
F(t)={\rm Tr}[\zeta V(0) \zeta W(t)\zeta V(0) \zeta W(t)]\, , 
\end{equation}
which is the one investigated in the main MSS argument \cite{Maldacena:2015waa}.
Indeed,
the relation between  Eq.\eqref{new} and Eq.\eqref{coc}
is 
\begin{equation}
\label{rel}
\tilde{C}(t)=-{\rm Tr}[\zeta^{2}[W(t),V]\zeta^{2}[W(t),V]]={\rm Tr}[\zeta^{2}W(t)V\zeta^{2}VW(t)]+{\rm Tr}[\zeta^{2}VW(t)\zeta^{2}W(t)V]-F(t+i\beta/4)-F(t-i\beta/4)\, ,
\end{equation}
with $V\equiv V(0)$. 
The first two terms can be re-absorbed in thermal state normalizations, while only the last $F$-terms contribute to the time growth \cite{Maldacena:2015waa}. 

As mentioned above, for $t_{d}<<t<<t_{*}$, the $F(t)$ is expected to be approximately 
factorable to
\begin{equation}
\label{Fdd}
F(t)\simeq F_{d}\equiv {\rm Tr}[\zeta^{2}V\zeta^{2}V]{\rm Tr}[\zeta^{2}W(t)\zeta^{2}W(t)],
\end{equation}
as a product of disconnected correlators of $V$ and $W$. 
In the large N-field limit, the factorization has the form 
\begin{equation}
\label{laag}
F(t)=({\rm Tr}[V\zeta W(t)\zeta ^{3}])^{2}+({\rm Tr}[W(t)\zeta V \zeta^{3}])^{2}+{\rm Tr}[\zeta^{2}V\zeta^{2}V][\zeta^{2}W(t)\zeta^{2}W(t)]+O(N^{-2})
\end{equation}
where the 
first two terms exponentially decay for $t>>t_{d}$. 

\vspace{0.1cm}

In a chaotic system, the $F(t)$  departs from the constant value $F_{d}$ 
after a critical scrambling time $t_{*}$. 

MSS conjectured 
that such a growth rate is bounded as follows: 
\begin{equation}
\label{rate}
\frac{d}{dt}\Delta F(t)\leq \frac{2\pi}{\beta}\Delta F(t)\rightarrow \Delta F=\delta \,  {\rm exp}\,\lambda t\, , 
\end{equation}
where $\Delta F(t)\equiv F_{d}-F(t)$, with $f=F(0)$ is an initial small quantity, and 
\begin{equation}
\label{lam}
\lambda \leq 2\pi\beta^{-1}\, . 
\end{equation}
The conjecture is supported by a simple argument that we review in Appendix A. 
The main assumption of the MSS argument consists into the
 treatment of the error variable $\epsilon$ as 
\begin{equation}
\label{ddF}
\frac{d}{dt}(F_{d}-F(t))\leq \frac{2\pi}{\beta}(F_{d}-F(t)+\epsilon)
\end{equation}
where the error $\epsilon\equiv \epsilon_{t}$ depends on the reference time considered.
In other words, Eq.\eqref{ddF} converges to Eq.\eqref{rate} only under proper assumptions
for the time $t$ that imply $\epsilon_{t}<<1$. In case GR, it is sufficient 
to consider a time scale much below the BH information scrambling time
for ensuring that Eq.\eqref{ddF} converges to Eq.\eqref{rate}, i.e. $\epsilon_{t<<t_{*}}<<1$. 
Such an apparently innocuous assumption can be violated in  $f(R)$ gravity 
where metric instabilities with a new Lyapunov exponent $\lambda_{g}\geq \lambda$ are present. 
Indeed, the correlators can violate the Eq.\eqref{ddF} bound for $t<t_{*}$, if the characteristic 
time scale related to the metric instability is below the thermal MSS Lyapunov exponent. 
In this case, a new characteristic scrambling time, related to the metric instabilities, 
appears as $t_{g*}\sim \lambda_{g}^{-1}\log (\lambda_{BH}^{0})^{-1}$.   
In particular, for BH growing solutions, Eq.\eqref{ddF}  is replaced by 
\begin{equation}
\label{ddF1}
\frac{d}{dt}(F_{d}-F(t))\leq \lambda_{BH}(0)e^{\lambda_{g}t}(F_{d}-F(t)+\epsilon)\, ,
\end{equation}
leading to a double exponentialization of the $F$-function. 

Behind the MSS violation coming from $f(R)$ gravity metric instabilities, 
there is an important difference 
between $f(R)$ and GR  related to their BH thermodynamical properties. 
Indeed, an important assumption of the MSS conjecture is 
that the N-system has a fixed temperature $T$ and thus 
it is in thermal equilibrium. In other words, the chaos effects of correlators
are thought as a non-equilibrium transient before reaching an equilibrium 
to the final thermalization. Indeed the Hawking radiation is computed from semiclassical saddle approximations of the Euclidean 
quantum gravity action, coinciding with the partition function of the BH emitting Hawking radiation in thermal equilibrium 
with the external environment. 
On the other hand, if the BH metric 
has a horizon area growing without reaching any constant plateau in time, as an effect of growing metric instabilities of $f(R)$ gravity, 
then a perfect thermal equilibrium 
will be never reached. 
As pointed out  in Ref.\cite{Addazi:2016prb}, 
a dynamical BH horizon leads to  revisit the
standard BH thermodynamical considerations,
including the BH temperature.  
The time exponential growing of the BH radius 
leads to a continuous increasing of chaos in the system 
without reaching any asymptotic thermalization. 

In Ref.\cite{Maldacena:2015waa}, MSS  comment on the possible effect of higher-derivative corrections 
to Einstein gravity, arguing that the bound does not receive any corrections. 
Such a statement is based on a result 
holographically connected to Einstein gravity
related to  the AdS/CFT correspondence \cite{SS,Shenker:2014cwa}, that is:
\begin{equation}
\label{Ftt}
F(t)|_{t>>\beta}=f_{0}-\frac{f_{1}}{N^{2}}{\rm exp}\frac{2\pi}{\beta}t+O(N^{-4})\, , 
\end{equation}
where $f_{0,1}\sim O(1)$ are positive constants 
which depend on the particular $V,W$ considered. 
Such a result saturates the MSS bound, 
suggesting a $t_{d}\sim T^{-1}_{BH}$ and 
$t_{*}=T^{-1}_{BH}\log N^{2}$.  
However, the apparent disagreement of our results with CFT computations in the case of $f(R)$ metric instability can be understood
from several prospectives. First of all, $f(R)$ gravity cannot be fully identified with a
GR with quantum higher-derivative corrections. Indeed, $f(R)$ gravity can be rewritten 
as GR with a scalar field after a conformal transformation.  In this sense, the MSS bound stability with respect to higher derivatives does not apply on $f(R)$ gravity. 
On the other hand,
the dS/CFT holographic equivalence is notoriously a not 
so solid conjecture as (specific) AdS/CFT cases. 
Indeed, the holographic principle works  
 in physical systems where no any metric perturbations destabilized the fixed 
(A)dS background. In a dynamical background, the holographic principle may 
loose any controllability and reasonable sense. 

The case of a BH horizon saturating a de Sitter Hubble radius, the extremal 
Schwarzschild-de Sitter solution,  was never studied from the perspective of
dS/CFT and, in principle, such a correspondence may be lost in $f(R)$ gravity or 
just from quantum gravity higher-derivative operators beyond specific special and integrable cases such as\footnote{
For previous attempts to study Schwarzschild-de Sitter solutions in dS/CFT for a stable background see Refs. \cite{dSCFT1,dSCFT2,Addazi:2016wpz}.}  $AdS_{5}\times S_{5}/CFT_{4}, AdS_{2}/CFT_{1}, ...$.

\section{The Geodesic instability}

In classical chaotic systems, the Lyapunov parameters 
control the average rate of convergence or divergence 
of body trajectories within the phase space $\{X_{i},P_{i}\}$,
with $i$ the space-coordinate index. 

Let us consider the generic classical equation of motion 
\begin{equation}
\label{eqa}
\frac{dX_{j}}{dt}=P_{j}
\end{equation}
where 
$P_{j}\equiv P_{j}(X_{i})$.
Let us consider a certain trajectory 
$\bar{X}_{i}(t)$ which is a solution of the Eq.\eqref{eqa}.
In order to perform the geodesic stability analysis, 
we can consider the linearized equation of motion 
around the trajectory $\bar{X}_{i}$,
as follows: 
\begin{equation}
\label{eqb}
\frac{d\delta X_{j}(t)}{dt}=K_{ij}(t)\delta X_{j}(t)
\end{equation}
where $K_{ij}(t)=\partial P_{j}/\partial X_{i}|_{X_{j}(t)}$.
The integral solution of the linearized equation can be rewritten in the formal form 
\begin{equation}
\label{eqc}
\delta X_{i}=U_{ij}(t)\delta X_{j}(0)
\end{equation}
where the  operator $U$ satisfies the following equation: 
\begin{equation}
\label{eqd}
\dot{U}_{ij}(t)=K_{im}U_{mj}(t)\, . 
\end{equation}
The principal Lyapunov parameter $\lambda$
is related to the following asymptotic limit: 
\begin{equation}
\label{eqe}
\lambda={\rm lim}_{t\rightarrow \infty}\frac{1}{t}{\rm log}\Big( \frac{U_{jj}(t)}{U_{jj}(0)}\Big)\, .
\end{equation}
However, in our case, the particle orbit has a radius 
that is thought, in first approximation, 
as slowly growing in time:
\begin{equation}
\label{eqe}
\lambda={\rm lim}_{t\rightarrow \infty}\frac{1}{t}{\rm log}\Big( \frac{U_{jj}(t)}{U_{jj}(0)}\Big)\sim \lambda_{0} \,e^{\lambda_{g}t}\, ,
\end{equation}
In the case of $\lambda_{g}>0$ and $t<<\lambda_{g}^{-1},$ such a dynamics would correspond to 
a spiral trajectory slowly increasing the radial coordinate in time. 
In the opposite regime, where $t>>\lambda_{g}^{-1}$, the radial instabilities 
would dominate on any other circular instabilities. 

In the case of circular geodetic orbits, 
the $K$ matrix and the Lyapunov parameter 
can be related to the metric 
tensor components:
\begin{equation}
\label{eqf}
\lambda=\pm \sqrt{K_{0}K_{1}}
\end{equation}
where 
\begin{equation}
\label{eqg}
K_{0}=\frac{d}{dr}\Big(\dot{t}^{-1}\frac{\delta \mathcal{L}}{\delta r} \Big),\,\,\, K_{1}=-(\dot{t}g_{rr})^{-1}\,.
\end{equation}
In our case, in the approximation of a slow radial time-variation, 
a pure circular trajectory will be substituted by a spiral trajectory with a slowly increasing radius in time. 
As a specific case of Eq.\eqref{eqe}, we obtain an exponential growth in time of the Lyapunov exponent: 
\begin{equation}
\label{eqh}
\lambda(t)=\sqrt{\frac{V_{r}''}{2\dot{t}^{2}}}\sim c\, e^{\lambda_{g}t},\,\,\, t<<\lambda_{g}^{-1}\, ,
\end{equation}
where
\begin{equation}
\label{eqg}
ds^{2}=f(r,r_{BH}(t))dt^{2}-\frac{1}{g(r,r_{BH}(t))}dr^{2}-r^{2}d\Omega^{2}\, ,
\end{equation}
which can be reduced to a point-like Lagrangian
\begin{equation}
\label{eqh}
2\mathcal{L}\simeq f(r,r_{BH}(t))\dot{t}^{2}-\frac{1}{g(r,r_{BH}(t))}\dot{r}^{2}-r^{2}\dot{\phi}^{2}\, , 
\end{equation}
and, from the Euler-Lagrange equations, we get the effective potential
\begin{equation}
\label{eqi}
V''_{r}=-2\frac{g}{f}\left(\frac{3ff'/r-2f'^{2}+ff''}{2f-rf'}\right)\, . 
\end{equation}
The Lyapunov exponent can be related to the effective potential as follows: 
\begin{equation}
\label{eql}
\lambda(t)=\frac{1}{2}\sqrt{(2f-rf')V_{r}''}\,. 
\end{equation}
In Eqs.\eqref{eqh}-\eqref{eql} a slow time variation has been considered, neglecting all time derivative
terms of the BH radius, i.e.  $\dot{r}_{BH}\simeq 0,\ddot{r}_{BH}\simeq 0,...$. 
With these onsiderations in mind, let us discuss now the quasi normal modes.


\section{The Quasi Normal Modes}

Let us consider a probe scalar field around the BH:
it is described by the Klein-Gordon (K.G.) equation in the curved background as 
\begin{equation}
\label{aka}
\Box \Phi \equiv \frac{1}{\sqrt{-g}}\partial_{\mu}(\sqrt{-g}g^{\mu\nu}\partial_{\nu}\Phi)=0\, , 
\end{equation}
where $\partial_{\mu}$ is the partial four-derivative. 

In a spherical space-time background, the K.G. 
equation has a solution that can be decomposed as 
\begin{equation}
\label{papp}
\Phi(t,r,\theta,\phi)=\frac{1}{r}R(r,t)Y_{lm}(\theta,\phi)\, , 
\end{equation}
where $Y_{l}\equiv Y_{l}(\theta,\phi)$ are the spherical harmonic function
with $l$ the angular momentum eigenvalues satisfying 
the following
\begin{equation}
\label{akakll}
\nabla_{\theta,\phi}^{2}Y_{l}^{m}(\theta,\phi)=\Big[\frac{1}{\sin \theta}\partial_{\theta}(\sin \theta \partial_{\theta})+\frac{1}{\sin^{2} \theta}\partial_{\phi\phi}^{2}\Big]Y_{l}^{m}(\theta,\phi)=-l(l+1)Y_{l}^{m}(\theta,\phi)\, , 
\end{equation}

inserting Eq.\eqref{papp} in the K.G. equation, 
one obtains the following equation for the radial part of the scalar field:
\begin{equation}
\label{KG}
\Big[ \frac{\partial}{\partial t^{2}}-\frac{\partial^{2}}{\partial r_{*}^{2}}+V_{s}(r)\Big]R(r,t)=0\, , 
\end{equation}
 known as the Regge-Wheeler wave equation \cite{RW}.
Let us  introduce now the tortoise radial coordinate 
\begin{equation}
\label{rss}
r_{*}=\int_{-\infty}^{\infty} \frac{dr}{f(r)},\,\,\, {\it for}\,\,\, r>r_{+}\, . 
\end{equation}
The $V_{s}(r)$ appearing in Eq.\eqref{KG} is the effective potential 
reading 
\begin{equation}
\label{VVA}
V_{s}(r)=f(r)\Big[\frac{l(l+1)}{r^{2}}+\frac{1}{r}\frac{df(r)}{dr} \Big]\, . 
\end{equation}
In the case of a generic integer spin field such a potential can be generalized as
\begin{equation}
\label{soi}
V_{s}(r)=f(r)\Big[\frac{l(l+1)}{r^{2}}+(1-s^{2})\frac{2M}{r^{3}}+(1-s)\Big(\frac{1}{r}\frac{df(r)}{dr}-\frac{2M}{2r^{3}}\Big)\Big]\, . 
\end{equation}
\begin{figure}[ht]
\centerline{ \includegraphics [width=0.5\columnwidth]{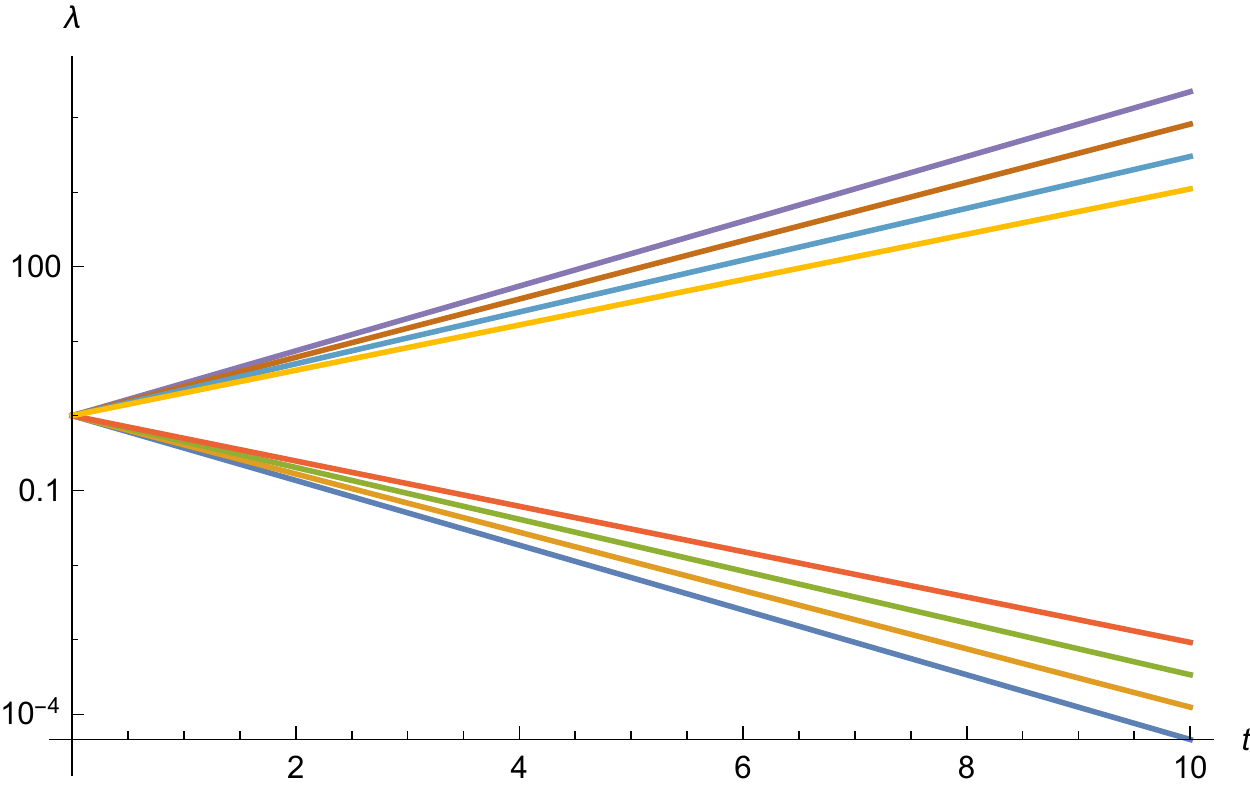}}
\caption{The photon-sphere Lyapunov parameters as a function of time for evaporating and antievaporating solution branches. }
\end{figure}
However, in case of metric instabilities, the background has a radius changing in time;
thus the potential $V_{s}(r)$ would have an effective time dependence through the time variation of the BH radius.
Assuming that the time-variation of the BH radius is  much slower than the characteristic field frequency, we can neglect any $\dot{M},\ddot{M},...$ derivatives and 
we can treat the 
time dependence inside the potential 
redefining $f(r)$ and $M(r)$ as 
\begin{equation}
\label{red}
M\rightarrow M(t),\,\,\,\, f(r)\rightarrow f(r,r_{BH}(t))=1-\frac{r_{BH}(t)}{r}\equiv 1-\frac{2M(t)}{r}\, ,  
\end{equation}
\begin{equation}
\label{jjja}
V_{s}(r,t)\simeq f(r,r_{BH}(t))\Big[\frac{l(l+1)}{r^{2}}+(1-s^{2})\frac{2M(t)}{r^{3}}+(1-s)\Big(\frac{1}{r}\frac{df(r,r_{BH}(t))}{dr}-\frac{2M(t)}{2r^{3}}\Big)\Big]\, ,
\end{equation}
where time derivative terms of $r_{BH}(t)$ are neglected. 
The usual approach for solving the Regge-Wheeler equation is a separation of the wave function 
assuming $R(r,t)=\Psi(r)\zeta(t)$, with $\zeta(t)\sim e^{-i\omega t}$, obtaining 
\begin{equation}
\label{sep}
\frac{d^{2}\Psi(r)}{dr_{*}^{2}}+[\omega^{2}-V_{s}(r)]\Psi(r)=0\, , 
\end{equation}
where $\omega$ corresponds to the QNM frequency, 
having a real and an imaginary parts
as 
\begin{equation}
\label{kaaaa}
\omega=\omega_{R}+i\omega_{I},\,\,\,\, \omega_{R}\equiv {\rm Re}\omega,\,\,\,\omega_{I}={\rm Im}\omega\, ,
\end{equation} 
with $\omega_{R}$ the real oscillation frequency and $\omega_{I}$ 
 the imaginary part related to damping or exponential instabilities 
of the field modes. 
However, the separation of radial and  time parts cannot be performed, 
in our case,  if the background time variation is comparable 
with the inverse of the characteristic real and imaginary parts.

Assuming that 
$m<< \omega_{R},\omega_{I}$, the time derivative can be neglected and we obtain 
\begin{equation}
\label{sep}
\frac{d^{2}\psi(r,t)}{dr_{*}^{2}}+[\omega(t)^{2}-V_{s}(r,t)]\psi(r,t)=0\, . 
\end{equation}
where $\psi$ has a slow varying time variation
and $R(r,t)=\psi(r,t)e^{-i \omega(t) t}$, where $\omega(t)$ derivatives are neglected, i.e. $d^{n}\omega/dt^{n}=0$ for any $n>0$. 
Under this assumption, we can search for QNM 
as saddle solutions of the following WKB condition:
\begin{equation}
\label{aasdff}
\frac{i(\omega^{2}(t)-V(r_{0}(t)))}{\sqrt{-2V''(r_{0}(t))}}+{\rm const}=n+\frac{1}{2}\, , 
\end{equation}
where the only time dependence in the potential enters through the $r_{0}(t)$.
This implies that $\omega$ has a slow time dependence as $\omega \equiv \omega(t)$. 
The $V''(r_{0}(t))$ is the second derivative of the effective potential with respect 
to $r$ at the maximum point $r_{0}$ related to $dV/dr_{*}|_{r_{*}=r_{0}}=0$. 
In our case, the maximum is slowly changing in time. 
In the deep WKB regime, $l>>1$ and 
\begin{equation}
\label{kQNMM}
\frac{d^{2}}{dr_{*}^{2}}\Psi+q\Psi=0,\, , 
\end{equation}
\begin{equation}
\label{kQNMM2}
\frac{q(r,t)}{2\frac{d^{2}q(r,t)}{dr_{*}^{2}}}=i(n+1/2)\,, 
\end{equation}
and 
\begin{equation}
\label{kQNMM3}
q(t,r_{0}(t))\simeq \omega^{2}(t)-f(r,r_{BH}(t))\frac{l^{2}}{r^{2}}\, . 
\end{equation}

Such a condition corresponds to the following QNM frequency: 
\begin{equation}
\label{omegaQNM}
\omega_{QNM}(t)=l\sqrt{\frac{f_{0}}{r_{0}^{2}}}-i\frac{(n+1/2)}{\sqrt{2}}\sqrt{-\frac{r_{0}^{2}}{f_{0}}\Big( \frac{d^{2}}{dr_{*}^{2}(f/r^{2})}\Big)\Big|_{r=r_{0}(t)}}\, . 
\end{equation}
which can be rewriten as 
\begin{equation}
\label{oQN}
\omega_{QNM}(t)=\Omega_{c}(r_{0}(t))l-i(n+1/2)|\lambda(t)|\, . 
\end{equation}

In the case of the near Schwarzschild-de Sitter BH, 
$\lambda=\Omega_{c}$
\begin{equation}
\label{omam}
\omega_{QNM}=\Omega_{c}(r_{0}(t))[l-i(n+1/2)]\, . 
\end{equation}
where 
\begin{equation}
\label{omegaQNMdS}
\Omega_{c}(r_{0}(t))=\frac{r_{\Lambda}(t)-r_{+}(t)}{2r_{+}^{2}(t)}\simeq \epsilon \frac{1}{2}e^{-a_{\pm}t} \, .
\end{equation}
with $\epsilon=r_{\Lambda}(0)-r_{+}(0)<<r_{+}(0)$. 

In the case of a perfectly extremal Schwarzschild-de Sitter BH, the horizons coincide as soon as $r_{\Lambda}=r_{+}$ 
and thus $\Omega_{c}=0$; i.e. no  QNM is found.  From Eq.\eqref{omegaQNMdS}, one expects that $\Omega_{c}$ decreases with the cosmological time if $a_{+}>0$ (antievaporation) while increases for other cases (evaporation). 
This corresponds to the  case where the QNMs can be altered by the slow dynamical 
variation of the BH geometry sourced by  gravitational actions like $f(R)$ gravity. 


\section{The Anomalous Black Hole Shadow}

As mentioned  in the Introduction, 
the BH shadow is defined as the area 
delimited by the critical photon-trajectory\footnote{Here, we will always consider radial hierarchies in the case of spherically symmetric BHs; for the Kerr case, such radii remain the same with good approximation \cite{Shadow1,Shadow2,Shadow3}. }, $r_{C}= 3r_{BH}/2=3M_{BH}$
\cite{classic1}.
Nevertheless, the critical impact parameter, where 
light is trapped, is higher than the $r_{C}$ by a factor $\sqrt{3}$, 
i.e. $b_{C}\simeq \sqrt{3}r_{C}\simeq 5.2\, r_{BH}$ \cite{classic2}. 
The photon-sphere appears as a luminous ring around the BH 
shadow, notoriously interpreted as the evidence of the BH horizon.
Indeed, EHT observed the $M87^{*}$ center with a BH size resolution,
measuring the photon-ring around the BH shadow.
The interpretation favors the presence of a BH horizon 
and it seems to rule out horizonless exotic compact objects \cite{EHT1,EHT2,EHT3}
\footnote{See also Refs.\cite{Bambi:2019tjh,Vagnozzi:2019apd,Banerjee:2019nnj,Li:2019lsm,Khodadi:2020jij}
for tests of BH hairs, extra dimensions and BH spin 
from $M87^{*}$.}.

Here, we wish to remark that a test for the BH horizon existence 
does not coincide with a general test for Extended Theories of Gravity,
including $f(R)$ gravity. Indeed, there are several BH solutions, 
having an external null-like event horizon, obtained in  large classes  of alternative theories of gravity and, in particular, in  
$f(R)$ gravity. In this category can be included the
 Schwarzschild-de Sitter  solutions discussed above. 
Nevertheless, the metric instabilities of $f(R)$ gravity, related to  
 the new Lyapunov exponent $\lambda_{g}$, 
can give rise to  anomalous variations of the BH radius and, thus,
of the the BH shadow with respect to similar features in GR.  

Specifically, the ratios among the critical photon-orbit,
the impact parameter and the photon-ring are fixed 
and controlled by the time-varying BH radius:
\begin{equation}
\label{hierarchy}
R_{+}=R_{-}+\Delta R\simeq kb_{C}+\Delta R\simeq \sqrt{3}kr_{C}+\Delta R= \sqrt{3}(3/2)kr_{BH}(t)+\Delta R=\sqrt{3}(3/2)kr_{BH}(0)e^{\lambda_{g} t}+\Delta R\, ,
\end{equation}
where $R_{\pm}$ are the maximal and minimal radii delimiting the photon-ring and $\Delta R$ is the photon-ring thick length;
$k$ is a parameter giving the numerical hierarchy between $b_{C}$ and $R_{-}$ that will be discussed soon. 
Eq.\eqref{hierarchy} implies that the photon-ring is time-dependent and it is controlled by the metric Lyapunov exponent $\lambda_{g}$, while $\Delta R$ remains constant in time. 

For fixing the ideas, let us consider light emission with an intensity peak $I_{0}$ around  the critical photon-orbit,
as displayed in the left side of Fig.3. It is worth to remark that the emitted intensity profile  significantly differs 
from the observed intensity on terrestrial experiments. 
Indeed, in the right side of Fig.3, we show how the profile of emitted intensity would appear for an asymptotic observer. 
In this case,  the observed intensity peak appears as a narrow distribution around 
 $r/M\simeq 5.5$, while it is highly suppressed for $r/M<4$ and $r/M>6$ \cite{Shadow1,Shadow2,Shadow3}. 
The lensing ring emission is within a thin ring because of the
 BH demagnification effect at $r/M>6$ and the efficient gravitational captures within the BH shadow from $r/M<b_{C}$. 
 
Let us now consider the case of a positive geometric Lyapunov exponent:
 in this case, while the emitted intensity spectrum retains the same profile in time, 
the photon-ring dynamically grows in time preserving its geometrical profile (see Fig.3). 
The photon-ring, at a fixed time $t_{0}$, is absorbed by the growing BH shadow at $t_{1}>t_{0}$,
but it is replaced by higher distance photons within $R_{\pm}(t_{1}) >R_{\pm}(t_{0})$.  

\begin{figure}[ht]
\centerline{ \includegraphics [width=0.8\columnwidth]{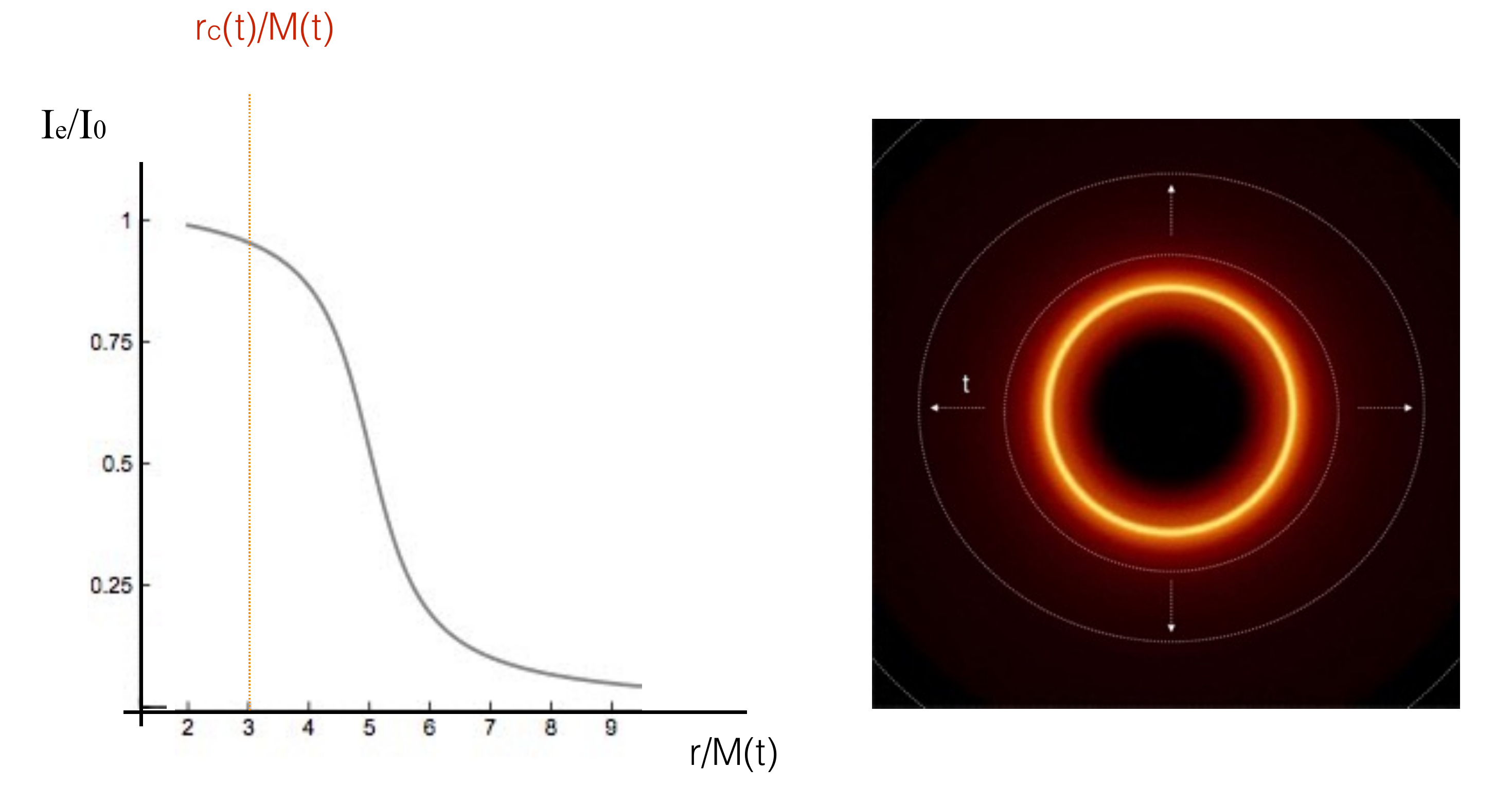}}
\caption{An example of emitted intensity profile $I_{e}$, normalized with the maximal peak $I_{0}$,
as a function of the radius, normalized by the time-dependent BH mass $M(t)$, is displayed on the left.
On the right side, we show its corresponding photon-ring intensity seen by an asymptotically far observer.
In case of $\lambda_{g}>0$, the photon-ring dynamically increases with time (here the case of $t>\lambda_{g}^{-1}$ is qualitatively displayed). }
\end{figure}

Such a phenomenon could be tested from BH shadow measurements,
searching for anomalous growing or shrinking of the 
photon-ring in time. Clearly, due to the very fine  measurements required for this pupose, it could be feasible for the Next Generation Event Horizon Telescope Design Program \cite{next}.

\section{Perspectives and Conclusions}

In this paper, we analyzed the emergence of chaos 
for circular trajectories around spherically symmetric BHs 
in the framework of $f(R)$ gravity. In particular, we focalized on the 
spherically symmetric solutions exhibiting an anomalous growing or shrinking of the BH 
event horizon area sourced by metric instabilities. 
We found that the trajectory instabilities are controlled by two Lyapunov exponents
as a double (time) exponentialization of perturbations. 
Our results provide examples of an explicit violation 
of the {\it Maldacena}-{\it Shenker}-{\it Stanford} (MSS) chaos bound.
The new extra Lyapunov parameter is related to the scalaron instabilities propagating in some classes of $f(R)$ gravity models. 
The new Lyapunov exponent of $f(R)$ gravity dominates on the MSS one
if the effective scalaron mass is heavier than a critical scale proportional to the BH temperature.
The BH temperature for supermassive BHs of the order of $10^9$ $M_{\odot}$ is around $10^{-17}$ Kelvin
corresponding to $10^{-21}\, eV$ or so. Thus if the scalaron has a mass 
higher than $10^{-21}\, eV$, the MSS can  be violated for a Supermassive BH in the context  of $f(R)$ gravity.  
We also showed that such metric instabilities lead to a distortion of  
the Quasi-Normal-Modes, with a potential  impact on  BH ringdown phase after the BH merging emitting gravitational waves. In conclusion, these features could be, from one hand,   signatures to test alternative theories of gravity. On the other hand, they could constitute straightforward explanations for classifying anomalies of compact objects, if revealed by next generation experiments.

\vspace{0.5cm}

{\bf Acknowledgements}
The work of A.A. is supported by the Talent Scientific Research Program of College of Physics, Sichuan University, Grant No.1082204112427. S.C. acknowledge the support of {\it Istituto Nazionale di Fisica Nucleare} (INFN) ({\it iniziative specifiche}  MOONLIGHT2 and  QGSKY). 
\vspace{0.4cm}

\section*{Appendix A: The Maldacena-Shenker-Stanford argument}

In this Appendix, we will complete the discussion of Sec.3, 
reviewing the MSS argument in favor of the chaos bound \cite{Maldacena:2015waa} in GR.  
First of all, let us consider the  analytic continuation of 
the $F(t)$ operator to complex times as follows:
\begin{equation}
\label{ana}
F(z=t+i\tau)=\frac{1}{Z}{\rm Tr}[e^{-(-\tau+\beta/4)H}Ve^{-(\tau+\beta/4)H}W(t)e^{-(-\tau+\beta/4)}Ve^{-(\tau+\beta/4)H}W(t)]\, . 
\end{equation}
In a finite volume and finite $N$, such a $F(z)$ is an analytic function 
in the $\tau\leq \beta/4$ range, while for $\tau=0$ the $F$ is always real. 

Let us define the function 
\begin{equation}
\label{fuf}
f(t)=\frac{F(t+t_{0})}{F_{d}+\epsilon}
\end{equation}
where $\epsilon$ is an infinitesimal quantity and $t_{0}$ a reference time. 

At this point the proof of the MSS bound will be obtained 
if the $f(t)$ satisfies the hypothesis of the following theorem.

\vspace{0.2cm}

{\bf Theorem.} {\it Let us consider a function $f(t)$ satisfying the following properties:

\vspace{0.1cm}

i) $f(t+i\tau)$ is an analytic function in $t>0$ and $-\beta/4\leq\tau\leq \beta/4$,
where $t,\tau$ are the real and the imaginary parts of the complex variable $z=t+i\tau$. 
Let us assume that $f(t)$ is real for $\tau=0$. 

\vspace{0.1cm}

ii) $|f(t+i\tau)|\leq 1$ in $t>0$ and $-\beta/4\leq\tau\leq \beta/4$.

\vspace{0.1cm}

Then, the (i)-(ii) hypothesis imply: 

\begin{equation}
\label{folll}
\frac{1}{1-f}\Big|\frac{df}{dt}\Big|\leq \frac{2\pi}{\beta}+O(e^{-4\pi t/\beta})\, . 
\end{equation}}

\vspace{0.1cm}

The Theorem can be proved considering a map of the half strip
$\{t>0, -\beta/4\leq\tau\leq \beta/4\}$ to the unit circle, in the complex plane. 
Such a transformation can be obtained using the following map:
\begin{equation}
\label{fofl}
Z=\frac{1-\sinh[2\pi \beta^{-1}(t+i\tau)]}{1+\sinh[2\pi \beta^{-1}(t+i\tau)]}\, . 
\end{equation}
The $f(Z)$ is an analytic function in the unit circle
as guaranteed by the (ii)-hypothesis. 
From the Schwarz-Pick theorem, the
function does not increase in the hyperbolic metric 
$ds^{2}=4dZ\bar{Z}/(1-Z\bar{Z})^{2}$, i.e.
\begin{equation}
\label{jaja}
\frac{|df|}{1-|f(Z)|^{2}}\leq \frac{|dZ|}{1-|Z|^{2}}\, . 
\end{equation}
Such an inequality corresponds to Eq.\eqref{folll}
for $\tau=0$ (having use the (i) hypothesis):
\begin{equation}
\label{fgh}
\frac{1}{1-f}\Big|\frac{df}{dt} \Big|\leq \frac{\pi}{\beta}\coth \Big(\frac{2\pi t}{\beta} \Big)(1+f)\leq \frac{2\pi}{\beta}+O(e^{-4\pi t/\beta})\, . 
\end{equation}

\vspace{0.2cm}

The rest of the MSS argument reduces to show that 
Eq.\eqref{fuf} satisfies 
$|f(t+i\tau)|\leq 1$ on three boundaries 
of the half strip $\{t>0,\,\,\, -\beta/4\leq \tau<\beta\}$
and that $f$ is constant in every points inside the half strip. 
Indeed, if the previous properties are proved, then the Phragm\'en-Lindel${\rm \ddot{o}}$f principle 
will guarantee that $|f|$ will be bounded by $1$ in every points of the
half-strip interior. 
On the $|\tau|=\beta/4$ edges, the $F$ reduces to 
\begin{equation}
\label{Ftbb}
F(t-i\beta/4)={\rm Tr}[\zeta^{2}VW(t)\zeta^{2}VW(t)]\, . 
\end{equation}
Such an equation can be considered as 
a contraction of two matrix elements 
$[\zeta VW\zeta]_{ij}$ and $[\zeta W V \zeta ]_{ij}$;
thus, we can apply the Cauchy-Schwarz inequality 
\begin{equation}
\label{aahll}
|F(t-i\beta/4)|\leq {\rm Tr}[\zeta^{2}W(t)V\zeta^{2}VW(t)]\, .
\end{equation}

For times larger than the dissipation timescale,
the operators factorize with a certain error 
\begin{equation}
\label{laal}
{\rm Tr}[\zeta^{2}W(t)V\zeta^{2}W(t)]\leq {\rm Tr}[\zeta^{2}W(t)\zeta^{2}W(t)]{\rm Tr}[\zeta^{2}V\zeta^{2}V]+\epsilon,
\end{equation}
where the error $\epsilon$ depends on the reference time $t$ and the first term on the right side is nothing but 
the definition of $F_{d}$. If we consider $t$ much minor than the scrambling time, then 
the $\epsilon$ will be small and the the $|f|\leq 1$ condition will be guaranteed. 

Let us now consider the boundary $t=0$, corresponding to $F(i\tau)$ with $-\pi/4\leq \tau \leq \pi/4$.
Following similar arguments as above, assuming $t$ sufficiently smaller than the scrambling time, 
it is easy to show that $|f|\leq 1$ is satisfied. 

Finally, to complete the proof, we show that $|f(z)|\leq C$ with $C$ as a constant with $C\geq 0$.
Using the chaos factorization and Hermiticity of the $V$ and $W$ operators, 
we obtain 
\begin{equation}
\label{FF}
|F(t+i\tau)|={\rm Tr}[\zeta^{1+\alpha}V\zeta^{1-\alpha}W\zeta^{1+\alpha}W\zeta^{1-\alpha}V]\sim {\rm Tr}[\zeta^{1+\alpha}V\zeta^{3-\alpha}V]{\rm Tr}[\zeta^{1+\alpha}W\zeta^{3-\alpha}W]\leq {\rm Tr}[\zeta V \zeta^{2}V]{\rm Tr}[\zeta W \zeta^{2}W]=F_{d}={\rm const}\, ,
\end{equation} 
where $\alpha=4\tau/\beta$. 
Thus, the theorem can be applied on Eq.\eqref{fuf} and 
we obtain 
\begin{equation}
\label{assum}
\frac{d}{dt}(F_{d}-F(t))\leq \frac{2\pi}{\beta}(F_{d}-F(t)+\epsilon)\, , 
\end{equation}
where $\epsilon<<1$ if $t<<t_{*}$, with $t_{*}$ the scrambling time. 
As we discussed in Sec.3, this last assumption should be relaxed in case of $f(R)$ gravity with metric instabilities,
evading the MSS argument.

\end{document}